\documentclass{sigchi}
\pdfoutput=1 

\toappear{Permission to make digital or hard copies of all or part of this work for personal or classroom use is granted without fee provided that copies are not made or distributed for profit or commercial advantage and that copies bear this notice and the full citation on the first page. Copyrights for components of this work owned by others than the author(s) must be honored. Abstracting with credit is permitted. To copy otherwise, or republish, to post on servers or to redistribute to lists, requires prior specific permission and/or a fee. Request permissions from Permissions@acm.org. \\
{\emph{CSCW'17}}, February 25--March 01, 2017, Portland, OR, USA. \\
Copyright is held by the owner/author(s). Publication rights licensed to ACM. \\
ACM 978-1-4503-4335-0/17/03. \$15.00  \\
DOI: http://dx.doi.org/10.1145/2998181.2998195}

\usepackage{balance}  
\usepackage{graphics} 
\usepackage{txfonts}
\usepackage{times}    
\usepackage[pdftex]{hyperref}
\usepackage{color}
\usepackage{textcomp}
\usepackage{booktabs, tabularx}
\usepackage{ccicons}
\usepackage{todonotes}
\usepackage{quoting}
\usepackage{verbatim}
\usepackage[export]{adjustbox}

\makeatletter
\def\url@leostyle{%
  \@ifundefined{selectfont}{\def\UrlFont{\sf}}{\def\UrlFont{\small\bf\ttfamily}}}
\makeatother
\def\pprw{8.5in}
\def\pprh{11in}

\definecolor{linkColor}{RGB}{6,125,233}
\hypersetup{%
  pdftitle={Mosaic: Designing Online Creative Communities for Sharing Works-in-Progress},
  pdfauthor={Joy Kim, Maneesh Agrawala, Michael S. Bernstein},
  pdfkeywords={Social computing, creativity, creative collaboration, art},
  bookmarksnumbered,
  pdfstartview={FitH},
  colorlinks,
  citecolor=black,
  filecolor=black,
  linkcolor=black,
  urlcolor=linkColor,
  breaklinks=true,
}

\begin{document}

\title{Mosaic: Designing Online Creative Communities for \\Sharing Works-in-Progress}

\numberofauthors{1}
\author{%
  \alignauthor{Joy Kim, Maneesh Agrawala, Michael S. Bernstein}\\
    \affaddr{Stanford University}\\
    \email{\{jojo0808, maneesh, msb\}@cs.stanford.edu}\\
}

\teaser{ 
\centering   
\includegraphics[width=\textwidth]{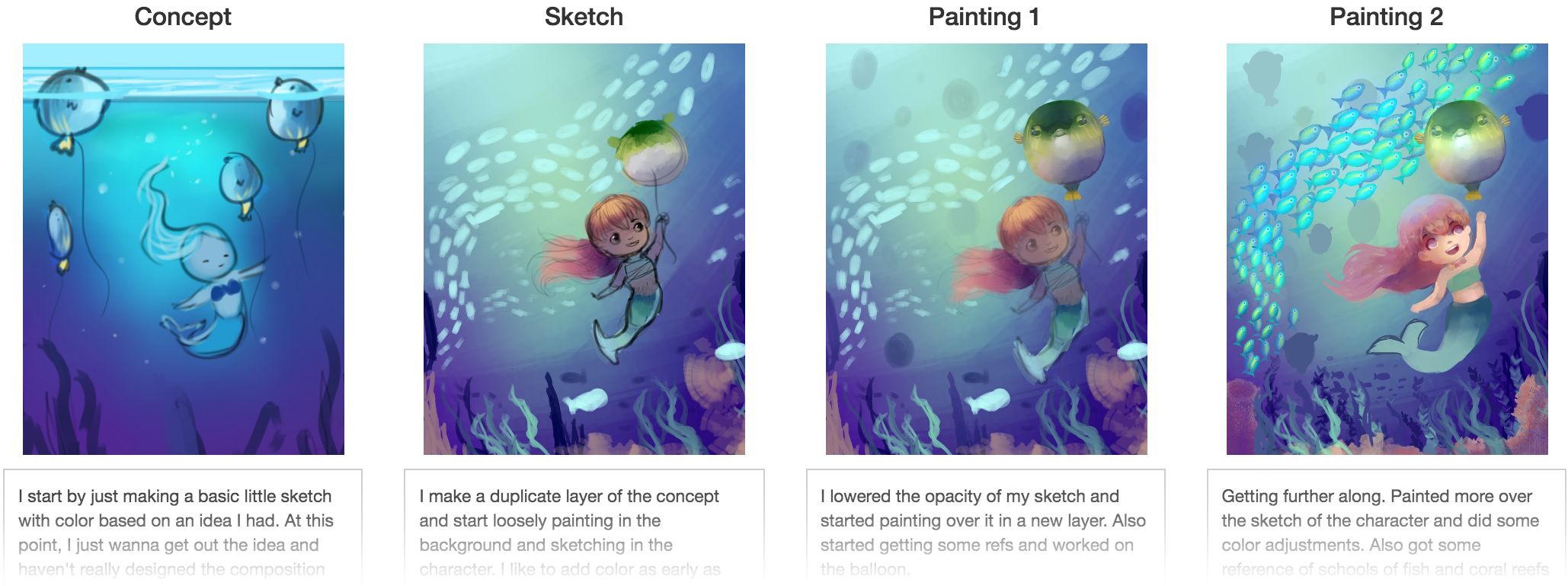}
\caption{Mosaic allows artists to share not just completed artwork but also their creative process. \emph{Fun Under The Sea} by masoto.}
\label{fig:main_figure} 
}
\maketitle

\begin{abstract}
Online creative communities allow creators to share their work with a large audience, maximizing opportunities to showcase their work and connect with fans and peers. However, sharing in-progress work can be technically and socially challenging in environments designed for sharing completed pieces. We propose an online creative community where sharing process, rather than showcasing outcomes, is the main method of sharing creative work. Based on this, we present Mosaic---an online community where illustrators share work-in-progress snapshots showing how an artwork was completed from start to finish. In an online deployment and observational study, artists used Mosaic as a vehicle for reflecting on how they can improve their own creative process, developed a social norm of detailed feedback, and became less apprehensive of sharing early versions of artwork. Through Mosaic, we argue that communities oriented around sharing creative process can create a collaborative environment that is beneficial for creative growth.
\end{abstract}


\category{H.5.3.}{Information Interfaces and Presentation
  (e.g. HCI)}{Group and Organization interfaces} \category{Collaborative computing}{}{}

\section{Introduction}
Online creative communities today focus on showcasing completed work, creating a climate where creators aim to produce work that is as impressive as possible to attract viewers and fans. 
In communities like those focused on art \cite{deviantart}, writing \cite{wattpad}, and design \cite{Behance}, a creator shares \emph{outcomes} by uploading finished pieces that are rewarded by views, favorites, or comments from others. The more views, favorites, and comments a submission gets, the more likely it is to appear in front of potential fans and other creators. Complementing these outcome-oriented communities, creators carve out \emph{process}-oriented spaces aimed at learning new techniques and receiving feedback from others, sharing in-progress work (e.g., \cite{artcrit}), creating and curating tutorials \cite{torrey2007pages}, or organizing events specifically for tackling creative challenges \cite{conceptartactivities}. 

But despite these efforts, creators encounter barriers to receiving thoughtful feedback in these online spaces. These barriers include the inability to tell which users are open to unsolicited feedback \cite{Marlow:2014:RAP:2531602.2531651} and a lack of mentors \cite{Guo:2015:CRO:2807442.2807469}. But most notably, simply posting work in the critique section of a creative community requires a creator to compete with finished work posted by others, discouraging the sharing of early work when feedback might be most useful. 
For example, on DeviantArt, users browse submissions by viewing single-image thumbnails (so that creators must optimize for views by creating single images that result in attractive previews); on /r/DestructiveReaders, a community centered around critiquing writing on reddit \cite{destructivereaders}, writers often ask for help by posting a link to their story (mirroring the way content is typically shared on reddit as a whole), omitting useful information such as their goals or what they have tried already; and on creative communities within Facebook and Tumblr, users feel they are spamming the community with unwanted content if they make multiple posts about the same creative work over time.


Instead, what if creative communities were designed to allow creators to share creative process as first-class content? Rather than just sharing finished work, creators could share in-progress snapshots of work to illustrate what they did and why.
Effective deliberate practice of a skill involves continually assessing one's creative process based on feedback and exploratory experiments~\cite{ericsson1993role,schon1983reflective}. Focusing on mastery~\cite{block1971mastery}, rather than performance, can increase self-perceptions of task-oriented confidence, especially for novices~\cite{Dow:2010:PPL:1879831.1879836}. Focusing on improving one's process can also have a significant effect on the quality of creative output: without engaging in a broad exploration of ideas, creators can experience fixation~\cite{jansson1991design}, but developing multiple ideas in parallel can produce a wider range of ideas and higher quality results~\cite{Dow:2010:PPL:1879831.1879836}.
By designing an environment that rewards sharing early work and clear explanations, instead of just rewarding good outcomes, we may create opportunities for creators to not only learn specific techniques from each other but also enable them to reflect more effectively on their own work.

In this paper, we focus on painting and illustration as an example of a domain especially dominated by outcome-oriented communities. To look at the types of interactions that arise when creators are instead able to focus on sharing process, we designed and launched Mosaic, an online social art platform where the primary method of sharing artwork is to upload multiple images illustrating the steps taken to complete it. By encouraging creators to show how their work develops over time, we enable an environment that values the communication of ideas and techniques.
We launched Mosaic and conducted an observational study in which 49 users created 76 Mosaic projects. These users successfully used work-in-progress steps from others to reflect on their own creative processes and wrote specific feedback for others. In addition, users expressed feeling less apprehensive about sharing early work on Mosaic compared to other creative communities they frequent, in part because Mosaic served as a social environment where it felt normal to do so. 

This paper suggests that building social systems for collaborative learning and growth require different social affordances than those developed for communities centered around sharing outcomes. It contributes online community design patterns and a system that demonstrates examples of such affordances.
Mosaic focuses specifically on illustration and art, but these design patterns may generalize to communities centered around many creative domains including music, film, writing, and design.
More broadly, we argue that planning, mistakes, experiments, techniques, and inspiration are normally hidden in social computing designs because they showcase finished work---but that these activities are valuable to communities where members may want to learn and support one another in their individual journeys of professional development.

\section{Related Work}
Mosaic's design draws from previous literature on the design of online creative communities as well as work studying creativity support for both novices and experts. In particular, it is inspired by existing practices for sharing information about creative process and how those practices support (and fall short of supporting) creators' goals.

\subsection{Online creative communities}

Creators who share a domain of interest often come together in communities of practice~\cite{Wasko:2005:WIS:2017245.2017249}; with online technology, creators from all over the world can build relationships with like-minded peers, learn new techniques, collaborate on projects, and work towards establishing their reputation in a community~\cite{nov2009motivational, Kuznetsov:2010:REA:1868914.1868950}. For example, a community might host contests or challenges where participating users create work based on the same theme, or provide social features such as messaging and forums that allow users to collaborate in co-producing work.
In addition, an online creative community giving feedback to each other can, in aggregate, provide positive mentoring experiences distinct from traditional offline mentoring~\cite{Campbell:2016:TPR:2818048.2819934}.
The interactions that users engage in on these communities may differ depending on whether users consider themselves professionals or hobbyists~\cite{Marlow:2014:RAP:2531602.2531651}.
Existing communities cover a wide range of interests, including songwriting~\cite{Settles:2013:LGT:2470654.2466266}, photography~\cite{nov2009motivational}, animation~\cite{Luther:2010:WWS:1880071.1880073}, and more. In this paper, we focus specifically on communities centered around painting and illustration. On these websites, users typically submit an image representing finished work (optionally accompanied by a short description), which allows them to build up a profile page that houses all of their submissions and acts as a portfolio of their activity.


The way these communities are designed has significant effects on how users understand who their audience is~\cite{marwick2011tweet} and how they interact and work with each other~\cite{Erickson:2000:STA:344949.345004}. For example, in online design communities, novices use signals of attention (e.g., likes) to determine which pieces of work to learn from and may tune their own sharing behavior to mimic strategies they see being used to share popular work~\cite{Marlow:2014:RAP:2531602.2531651}. As another example, interfaces that allow users to make judgments about the trustworthiness of others are essential for successful online collaborations~\cite{Luther:2010:WWS:1880071.1880073}. Leaders of collaborations, too, often bear a large burden to maintain group awareness, but interfaces can mitigate this responsibility by making group activity, signals of trust, and tasks to be completed concrete and transparent to the larger collaborating group~\cite{Luther:2013:RLO:2441776.2441891}. Models of successful creative processes~\cite{Settles:2013:LGT:2470654.2466266}---information that is normally invisible---could even be embedded in tools to encourage best practices, help creators find suitable collaborators, or help them figure out how to proceed in their work~\cite{Matejka:2009:CCR:1622176.1622214}.



Other work explores how larger crowds can come together to collaborate directly through the use of competitive marketplaces~\cite{99designs,threadless}, combining previous work by others~\cite{Yu:2011:CCC:1978942.1979147}, leader-generated constraints~\cite{Kim:2014:EEC:2531602.2531638}, remixing~\cite{Hill:2013:CCC:2441776.2441893}, training non-experts~\cite{Dontcheva:2014:CCL:2556288.2557217}, structuring the iterative feedback process~\cite{Xu:2015:CSU:2675133.2675140,Luther:2015:SAE:2675133.2675283}, and dynamically coordinating work by specialized experts~\cite{Retelny:2014:ECF:2642918.2647409}.
While we do not focus on direct collaboration in this paper, Mosaic builds on work that has shown how peer production can be improved through design and explores possible affordances for peer-supported learning and development. 
The design affordances Mosaic explores in the domain of sharing works-in-progress could be applied to crowd creativity work to enable more effective collaboration.



\subsection{The effect of the creative process on outcomes}

The process taken to create something can have a significant impact on creative outcomes; for example, prototyping several different designs for an advertisement in parallel (rather than iterating on a single design) results in better-performing ads as well as increased personal confidence for novices~\cite{Dow:2010:PPL:1879831.1879836}. These immediate effects on self-perception can also improve a creator's long-term ability to persevere~\cite{dweck2006mindset}. Conversely, a process where the creator chooses a design concept too early can result in design fixation~\cite{jansson1991design}, which can limit idea generation, even in experts~\cite{cross2004expertise}. Further complicating the creative process is the observation that a design problem can change as a creator explores solutions~\cite{schon1983reflective}, requiring the creator to be able to flexibly change their goals as they work.

Previous work has looked at specific interventions to the creative process to try and improve creative outcomes. Looking at examples can help an ideator expand their design space by allowing existing ideas to be combined and reinterpreted ~\cite{Herring:2009:GIU:1518701.1518717}, but only when examples have certain properties~\cite{chan2011benefits, Siangliulue:2015:TCI:2675133.2675239}. The timing of when examples appear in the creative process is also important; earlier tends to be better~\cite{Kulkarni2014}, and ideators that are presented with ideas when they are stuck present more ideas than those who are presented with examples at regular intevals. In fact, being presented with examples at regular intervals is worse than being presented with no examples at all~\cite{Siangliulue:2015:PTE:2757226.2757230}. Other work has looked at using a shared idea map to help groups generate diverse ideas collaboratively~\cite{Siangliulue:2015:TCI:2675133.2675239}, and even aiding in emulating specific expert strategies directly (such as by automatically generating drawing guidelines~\cite{Lee:2011:SRU:1964921.1964922}). This body of work shows how influencing the creative process can change creative outcomes, but it is still unclear how to incorporate these findings into creators' everyday practice. This can be especially difficult due to the fact that the process behind shared online work is often hidden. Mosaic, instead, attempts to complement this work by presenting a design for a social environment that helps creators focus on improving not just \emph{what} they produce but also \emph{how} they produce it.

Even without altering the creative process itself, simply reflecting on the creative process may help a creator think about new possible directions. Building a personal history using a timeline interface can provide a vehicle for identifying and reminiscing on key events~\cite{Thiry:2013:APH:2470654.2466215} and drive people to generate new interpretations of the past~\cite{Hodges:2006:SRM:2166283.2166294,Petrelli:2009:MHI:1518701.1518966}. We may see similar benefits among artists asked to document their practice through Mosaic. In addition, Mosaic users can reflect on their processes with others through the form of feedback, which may help them identify gaps between their intent and how others perceive their work~\cite{feldman1971varieties}. Those who help by participating in this reflective process can also benefit from newly generated insight~\cite{boud2014peer}.





\section{Formative Study}
To better understand the challenges that creators face in the creative communities they use to share their work, we conducted semi-structured interviews with ten intermediate-level creators (nine female, one male) recruited through posts in anime, video game, and comic fandom art communities on Facebook, Tumblr, and DeviantArt. Creators' ages ranged from 18 to 39 years old ($M = 27.4$), with occupations ranging from college student, full-time freelance illustrator, and QA developer. All participants had been or currently were active users of DeviantArt, and most additionally created posts about their art activity a few times a week on other social media platforms such as Facebook, Tumblr, Twitter, and Instagram.

Six of ten interview participants described their use of existing social media platforms for sharing art as oriented around exposure; they also described these platforms as not very useful for feedback, but use them anyway because they want to reach as many potential fans as possible.
Eight interview participants stated that they occasionally post single snapshots of in-progress work on online communities, but these serve mostly as a social update to engage those who follow them. Three participants mentioned that they had never documented their process in a step-by-step format at all, being unsure as to whether it was something their audience wanted or because they were not confident that they could successfully teach others. One participant mentioned being explicitly told to stop posting in a Facebook community after having uploaded several images about a project in a row.

Attempts at sharing the process behind their own artwork was met with various barriers, with four participants describing the interface design of these existing platforms as the main obstacle:

\begin{quoting}
\emph{You're trying to keep it in one post, but it's so much to keep track of... It was just, I guess, a lot of UIs and everything not really designed for that kind of thing where it's just...}

\emph{Then on DeviantArt, my god. Trying to get all the screencaps into one gigantic document was just ugh.}
---P8
\end{quoting}

Despite only being able to see the final outcome most of the time, the way interviewees viewed other artists' finished work on existing creative communities was in terms of process. Six participants said seeing good artwork fueled inspiration for them, but nine participants also explained that this was paired with a struggle (or even an inability) to demystify how the outcome they were seeing was achieved.


These interviews suggested that despite the popularity of creative communities online, and despite a desire to share and get feedback on process, many creators do not find the social and technical affordances of existing communities appropriate for process-oriented content. Exposing process was seen as helpful behavior that creators wanted to do but could not for technical or social reasons.
We address these needs in Mosaic, an online creative community where the main method of sharing work is to expose creative process.

\section{Mosaic}
We know that orienting learning and creative support around the creative process can result in benefits such as increased confidence and higher quality creative outcomes, but the design of online communities often presents barriers to creators who want to share information about their process. 

To explore potential designs for a community that enables social interactions oriented around creative process between artists, we created Mosaic: an online social platform where creators share artworks-in-progress. With the design of Mosaic, we envisioned a community where members are encouraged to share struggles in addition to successes, reflect on possible creative directions, and give and receive feedback informed by a creator's intent and goals for a piece of artwork. In this section, we describe how Mosaic allows creators to improve their own creative processes and those of others.

\subsection{Projects and Works-in-Progress}
In Mosaic, the main unit of content is called a \emph{work-in-progress} (Figure~\ref{fig:main_figure}). The work-in-progress is an image (either a photo or a screenshot) of a creative work that is not yet complete. This image is accompanied by a title and a short caption describing the image. For example, an artist starting work on an oil painting may create a work-in-progress representing their first step (e.g., drawing a sketch). This work-in-progress would include a description of any reasoning behind their step (e.g., why they chose a certain type of subject matter or how they chose a certain visual composition).

\begin{figure}[!t]
\centering
  \includegraphics[width=\columnwidth,cfbox=lightgray 0.5pt 0pt]{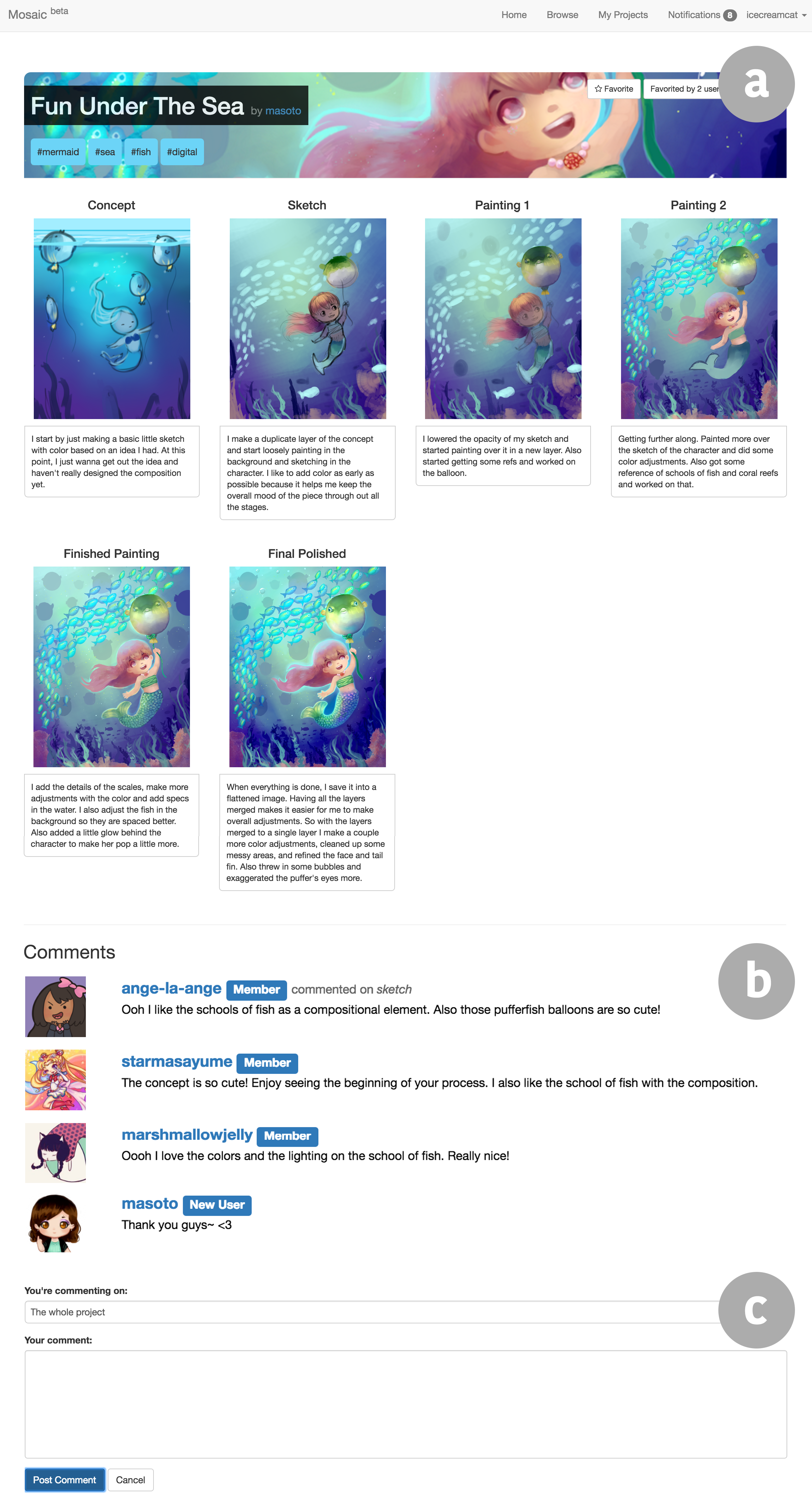}
  \caption{A Mosaic project. (a) The project consists of several snapshots of the artwork as it developed over time, including explanations by the artist describing what they did in each step and why. (b) Comments in Mosaic tend to be specific and considerate of the artist's creative intent. (c) Other users can comment on projects.}~\label{fig:project_figure}
\end{figure}

Creators can group works-in-progress in a \emph{project}  (Figure~\ref{fig:project_figure}), which represents a single creative work. That is, an artist working on a landscape painting may post a project representing that painting, adding works-in-progress representing stages of the piece as they go (e.g., Sketch, Mid-tone Wash, Blocking Shapes, Rendering Details). Optionally, creators can flag their project with a request for critique, which signals to other users that they are open to detailed feedback. Mosaic users can search works-in-progress using a search form that matches on text; if a user searches for ``sketch,'' they will be able to view all projects that contain a work-in-progress representing a sketch.

Social features enable artists to view what others are doing. The homepage consists of a feed of recent activity from the Mosaic community as a whole, showing comments, new projects, and updates to projects (that is, new works-in-progress that have been added to a project). Users are also able to follow other users and favorite projects so that they can be notified with an email when a user they follow creates new work or a project they have favorited is updated in any way. Lastly, users are able to comment on projects to share encouragement, feedback, or links to external resources.

\subsection{Scenario}
Making works-in-progress a first-class unit of shareable work normalizes a number of social interactions that are difficult on existing online creative communities. Below, we walk through a scenario illustrating some of the social advantages of using Mosaic to share creative work.

\subsubsection{Receiving helpful intermediate feedback}
Dawn is a novice artist who wants to be a professional illustrator. Though she has taken art classes through school in the past, she recently started taking artwork more seriously. Dawn joins Mosaic and, after being welcomed to the website, is prompted to upload a photo or screenshot of whatever piece of artwork she's currently working on. A few days ago, Dawn started work on small watercolor piece for a friend's birthday, so she creates a new \emph{project} titled ``Watercolor Gift'' and then adds a \emph{work-in-progress} to this project by taking a photo of her sketch so far and typing a short caption about her thought process behind the sketch.


\begin{figure}[!t]
\centering
  \includegraphics[width=\columnwidth,cfbox=lightgray 0.5pt 0pt]{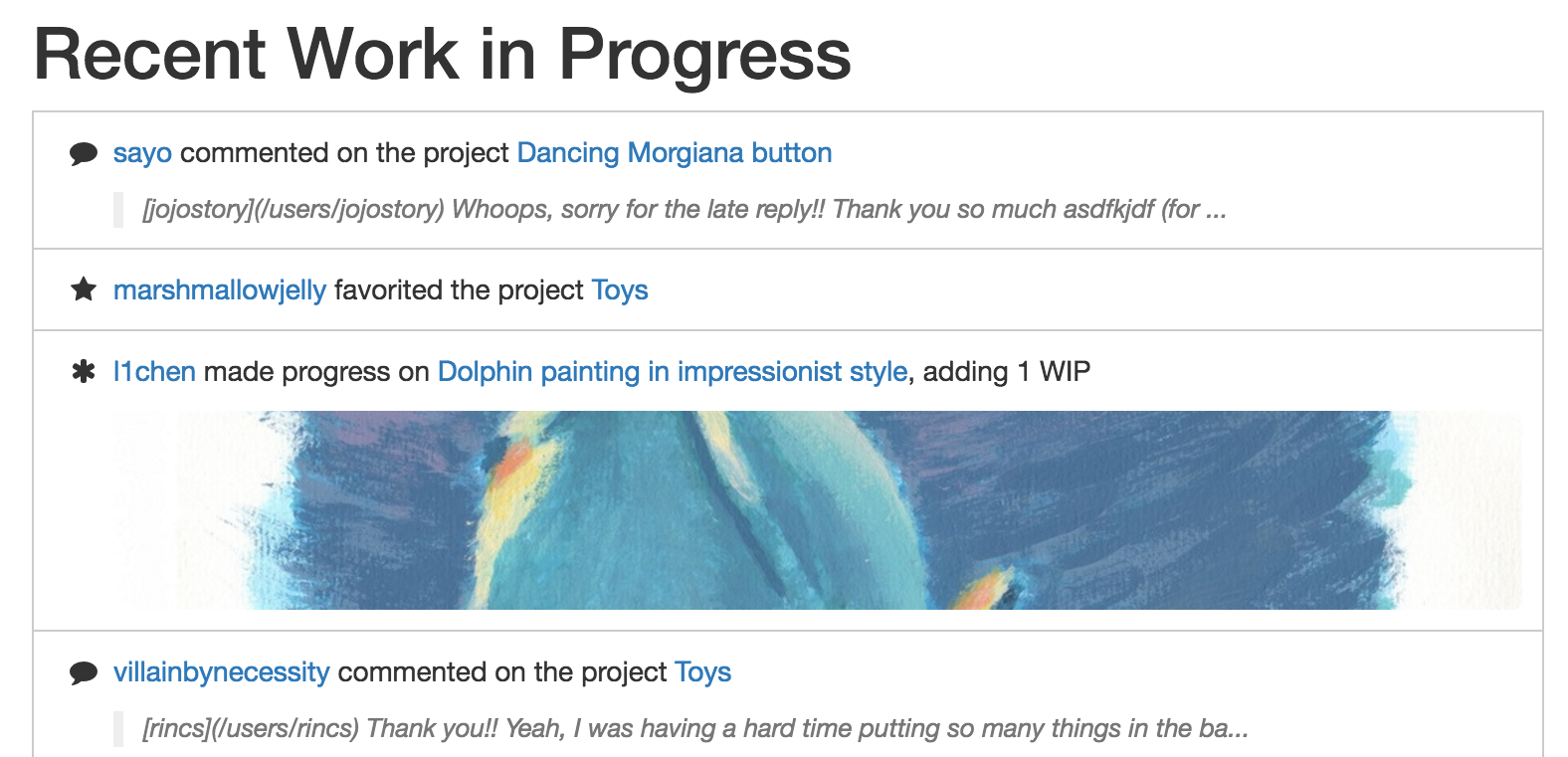}
  \caption{Users are shown a feed of recent activity when logged in. Activity in Mosaic is centered around progress made on projects rather than on finished artwork.}~\label{fig:recent_activity}
\end{figure}

\begin{figure}[!t]
\centering
  \includegraphics[width=\columnwidth,cfbox=lightgray 0.5pt 0pt]{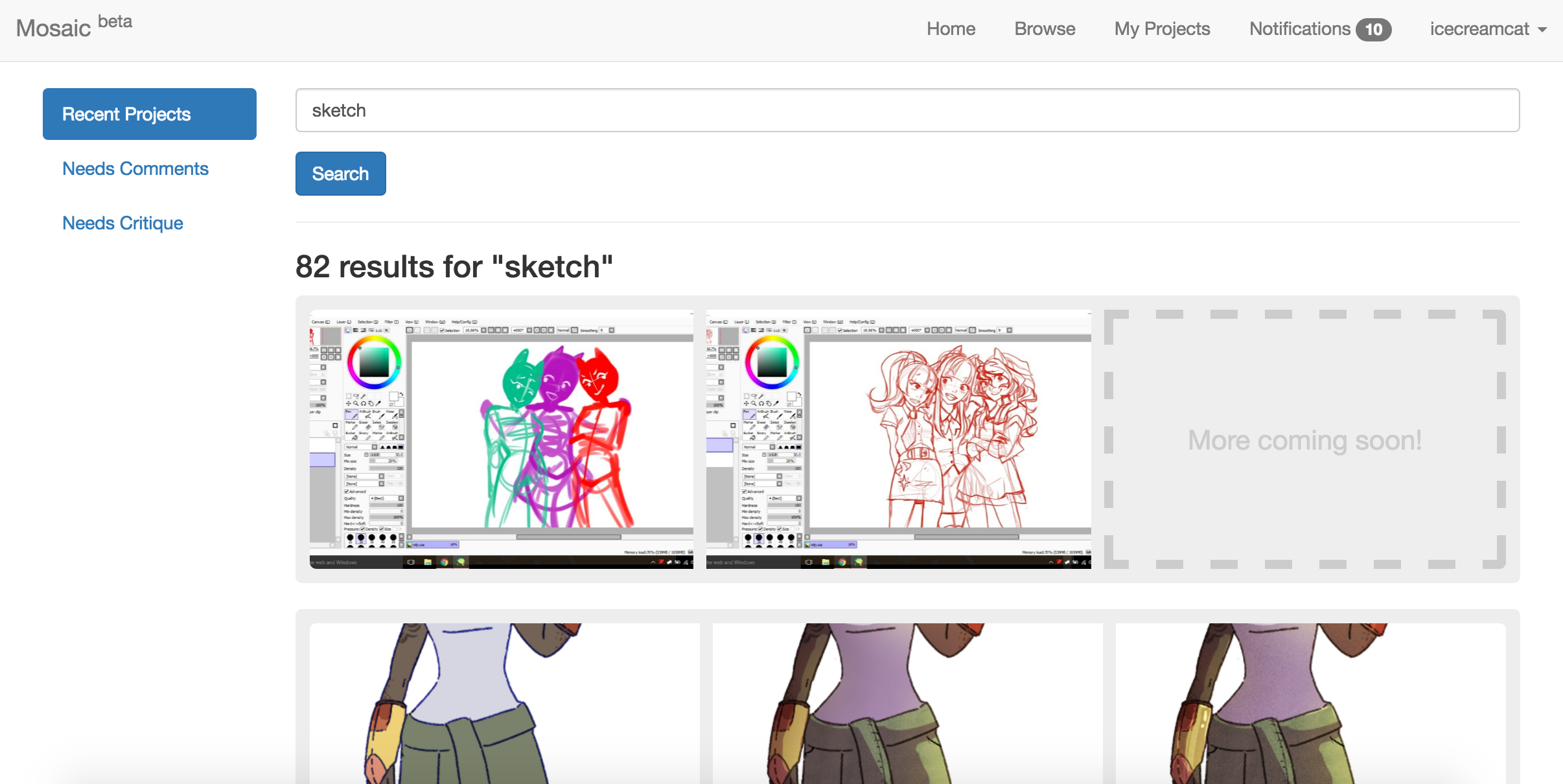}
  \caption{Users can search for projects by content or by techniques used.}~\label{fig:search_projects}
\end{figure}

Haruka is a freelance digital illustrator who is already a member of Mosaic. While browsing projects through the feed of recent activity seen on the Mosaic homepage (Figure~\ref{fig:recent_activity}), she sees a thumbnail of Dawn's project so far. The piece seems to be of a sketch of a person; Haruka has recently been studying anatomy and decides to takes a look to see if she can learn from this project. After clicking the project, she notices a mistake in the sketch, and leaves a comment.

Dawn receives an email notification about a new comment and logs into Mosaic. She realizes that Haruka is right about the mistake, and revises the sketch to address the issue. She takes a new photo and adds a new \emph{work-in-progress} to the existing \emph{project}, again accompanied by a caption summarizing the mistake and her solution. She leaves a comment responding to Haruka to thank her for the feedback, and makes a mental note to look out for similar anatomy mistakes in the future.

Later, after adding several more \emph{works-in-progress} photos to her project, Dawn finishes her watercolor piece. She posts the link to her Mosaic project on her social media accounts, noting that she receives a few likes and followers from posting content about her creative process.

\subsubsection{Learning new techniques}
Dawn is looking to start a new project and starts to browse Mosaic (along with other art community websites) for inspiration. In Mosaic, Dawn clicks on a few watercolor projects that seem visually similar to her own style, but notices from looking at the work-in-progress photos and captions that some of them are created with similar techniques used in a slightly different order. Others show a work-in-progress that shows the use of an additional technique that results in an unusual visual effect that Dawn has never tried before. Dawn feels motivated, and thinks about a project that would let her practice this new technique.

Dawn creates a sketch and uploads it as the first work-in-progress for a new project. However, while trying this new technique on scratch paper, Dawn finds she's having trouble getting it right. She takes a photo of these attempts and adds it as a new work-in-progress, noting in the accompanying caption that she's stuck. Dawn edits her project to flag it as requesting critique, which adds it to a special feed of projects that are occasionally emailed to users who choose to participate in giving feedback. Dawn later receives a clarifying comment from one of these users about the photo she uploaded, which lets her get unstuck.

\subsubsection{Secondhand learning}
Haruka is similarly struggling with a new technique for a digital painting she is working on. She normally creates illustrations in a cartoony style that makes use of solid colors and clean lineart; Haruka now wants to experiment with a more painterly look with her illustrations, but is having trouble figuring out how to do this efficiently. She uses the search function of Mosaic (Figure~\ref{fig:search_projects}) to filter for other projects that use that technique and look similar to her desired result. After finding a few suitable examples and examining their works-in-progresses, Haruka finds that she can use her old style of illustration until she is happy with the colors and lighting, then paint on top of this refine her illustration and hide lineart.

Haruka also knows that one of her favorite artists on Mosaic uses this technique in the same digital painting program as she does, so she leaves a comment on one of that artist's projects asking for more details about how this technique is done in that particular program. The artist later responds, and even updates the work-in-progress captions in their own project to address Haruka's question.

\subsection{A focus on process}
The above scenario highlights a key aspect of Mosaic's design: rather than focusing on showcasing final outcomes, Mosaic structures social interactions around units of content that show the process behind creative work. As a result, creators are able to directly ask for and receive help informed by the context of the creator's current skill level and their creative intent. The content shared on Mosaic is directly related to creators' goals to improve their process and learn new techniques.
In addition, Mosaic unlocks a number of interactions we know to be useful to creativity, such as being able to express intent \cite{feldman1971varieties}, receive feedback during the creative process \cite{Siangliulue:2015:PTE:2757226.2757230} rather than afterwards, reflect on progress \cite{schon1983reflective}, and determine when to reach out to others \cite{Erickson:2000:STA:344949.345004}. Mosaic can be visited at \url{http://www.artsaic.com}.

\section{Evaluation}
We designed Mosaic based on the hypothesis that sharing creative processes is difficult for creators because existing creative communities are designed to maximize the benefits of sharing creative outcomes. Instead, we propose an alternate design for a community designed around works-in-progress and seek to understand the types of interactions between creators that might result from such a design.
In this section, we report on an evaluation where we explored how this process of generating and sharing works-in-process strengthens creators' abilities to reflect and allows a community to generate more meaningful feedback and support for its members.


\subsection{Method}
In order to understand the difficulties artists face when seeking or sharing works-in-progress online, we conducted a field deployment of Mosaic to provide a meaningful example with which artists could compare and contrast against their past experiences. In contrast to a controlled study (which would require growing a control community without the draw of an established user base), a field deployment allowed us to prototype a design for a community for sharing process, to probe for existing practices surrounding works-in-progress and sharing knowledge, and to learn about ways in which Mosaic might disrupt or support these practices. This also allowed Mosaic users to compare their experiences on Mosaic with their current activity in communities they already frequent.

Over the course of four weeks, we launched Mosaic as an open beta and invited users from other hobbyist art communities to use Mosaic as a way to give and receive critique and as a platform for hosting and sharing in-progress work. To ensure that artists would be able to provide meaningful feedback to each other, we recruited from communities of artists with roughly similar backgrounds that had a wide range of skill levels (in this case, beginner to advanced intermediate artists from anime/video game/comic fandom communities from Facebook, Tumblr, and Deviantart). During this time, we logged all community activity, including the creation of projects, works-in-progress snapshots, project favorites, user follows, and comments. 

Because we wanted artists to create artwork they were personally invested in, we structured this study around the idea of creating a zine that would eventually be printed and advertised and sold to peers and fans in the community. Zines are typically small self-published anthologies of artwork created on a theme and, because of their self-published nature, are an accessible and popular way for creators of all skill levels to promote their work. In many online communities, they are often created as collections of fanart or fanfiction. On a practical level, zines provide opportunities for artists to meet each other and to cross-advertise their work. For this reason, we specifically recruited artists who worked in two-dimensional digital or traditional media. The zine will be compiled as a digital PDF and made available for download online. 

Artists who signed up on Mosaic were prompted to upload projects, which they could optionally submit to the zine. One submission per artist was allowed, though Mosaic users were able to create as many Mosaic projects as they liked. We allowed users to post both work they were currently working on as well as work-in-progress snapshots they may have had from previous work. Additionally, a peer voting round was used to select the artists that would be included in the zine, further incentivizing artists to do their best for the study task. The first 30 artists who made submissions to the zine were given a \$40 gift card.

We also conducted semi-structured interviews with the same ten artists who were interviewed during our formative study, all of whom were participants in the deployment.
Interview questions focused on their background and goals as an artist, the perceived benefits of sharing and viewing works-in-progress, the dynamics between themselves and other users on Mosaic, perception towards feedback both received and given, and attitudes toward the artwork they created during the study period. 

In order to understand how creators explained their own work and what motivated them to communicate with each other, we analyzed Mosaic comments and projects as well as the responses from our semi-structured interviews. First, to analyze Mosaic comments and projects, the first author generated codes by looking for recurring patterns in the text written by users for comments and works-in-progress. Using these codes, two researchers independently coded the same randomly selected subset of comments and works-in-progress (works-in-progress: $\kappa=0.64$; comments: $\kappa=0.65$) and discussed disagreements in codes. Code definitions were revised to resolve disagreements. The remaining dataset was split in half and separately coded by each researcher using the new codes. Second, we used a similarly inductive approach to develop themes in interview responses. These themes allowed us to understand the relationship between creating works-in-progress and sharing these snapshots with peers, how Mosaic's design might deter or encourage sharing information about creative process, motivations for creating works-in-progress and commenting on others works, and how their experience with the Mosaic community compared with their experience on existing online creative social platforms.

\section{Results}

\begin{figure*}[!t]
\centering
  \includegraphics[width=\textwidth,cfbox=lightgray 0.5pt 2pt]{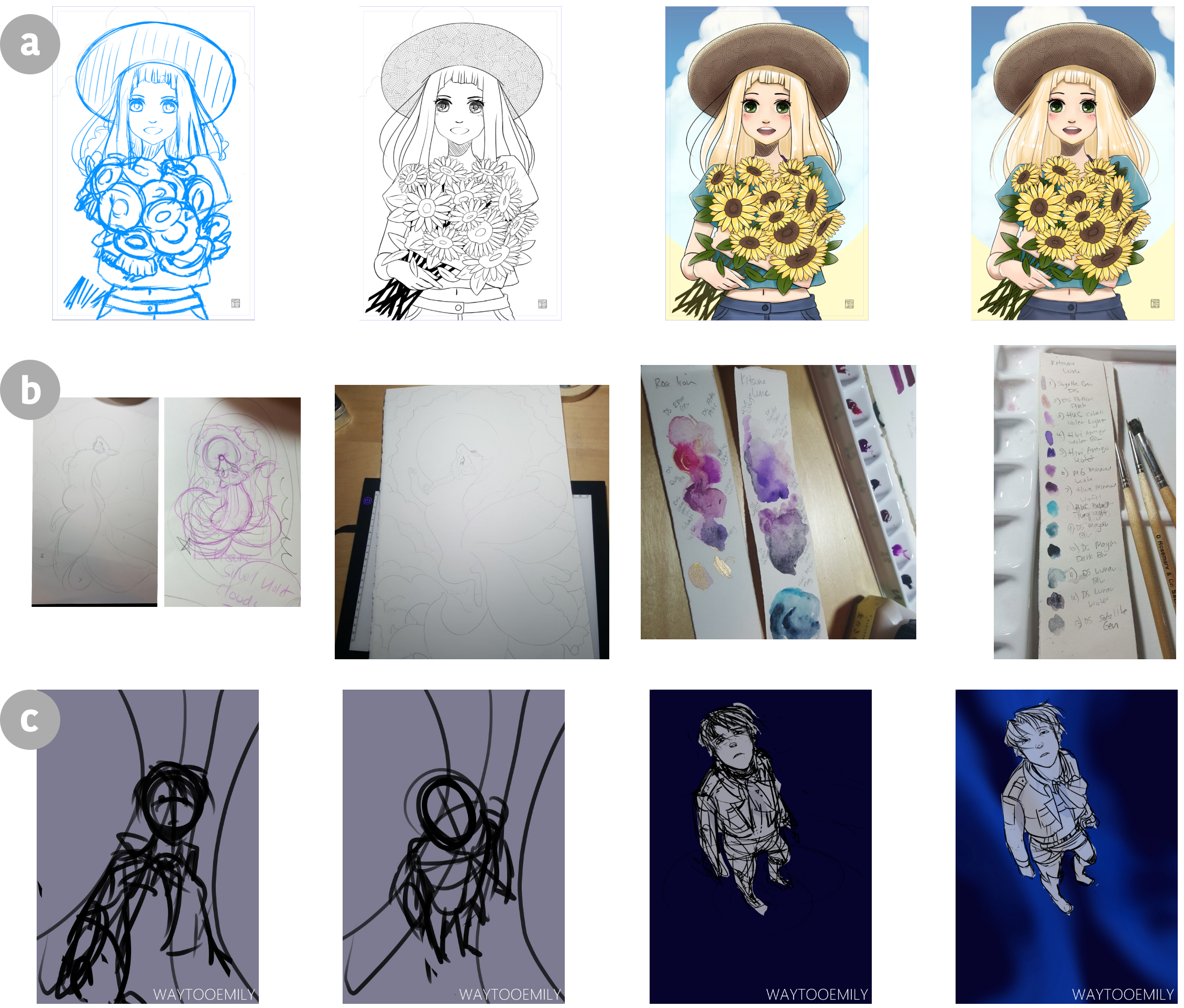}
  \caption{The first few works-in-progresses from some of the most popular projects on Mosaic. (a) \emph{Sunflowers} by marshmallowjelly. Many projects structured themselves around showing significant steps in the progress of the artwork. (b) \emph{Kitsune Lune} by starrydance. The creator shows an early sketch, as well as organizational tricks they use to remember what colors they use. (c) \emph{Like Satellites and Shooting Stars} by waytooemily. The creator shows a few early sketches illustrating how they chose a composition for the piece.}~\label{fig:projects_figure}
\end{figure*}

The projects uploaded to Mosaic allowed artists to compare their creative processes with each other, leading to both technical insights about how to improve as well as opportunities to validate their approaches to creative problems. Figure \ref{fig:projects_figure} shows works-in-progress from a few of the most viewed projects on Mosaic. These projects show the variety of types of information creators chose to share with others, including the both the ideation and technical steps behind an artwork.

During the study period, a total of 46 users created 69 projects, with projects containing an average of 5.26 ($SD=2.32$) works-in-progress.
Out of these, 38 projects were submitted as entries for the zine. During this time, 468 unique users made 1144 unique visits (3489 pageviews) to the Mosaic website, with about 40\% of incoming users arriving through links on existing social media sites like Facebook, Twitter, and Deviantart.


\subsection{Sharing process served as vehicles for reflection}

Table~\ref{tab:wip_categories} shows the various types of thoughts that creators documented while creating these works-in-progress, ranging from tutorial-like descriptions of steps taken during the creative process, to questions asked in the middle of the process, to explanations of the higher-level goal pursued by the creator.

Creators typically did not wait until finishing their artwork to post their works-in-progress, nor did they stop to upload works-in-progress as they worked. Instead, after saving images as they worked, artists would post one or more images representing substantial progress at the end of a working session (taking a median of 4.36 hours in between updates to their project) and reflect on their working session as a whole.


This perhaps explains why only 53\% of WIPs were objective descriptions of what was done in the artwork; these works-in-progress tended to focused more on loops of intent-attempt-assess and other internal thought processes than on the actual steps taken to acheive a visual effect:

\begin{quoting}
\emph{Initial sketch. The idea for this started as a quick doodle. It's been unseasonably warm this spring, so I've been wanting to draw something summery and I've been researching sunflowers for the garden so I've had these happy flowers on my mind recently...}

---Sketch, \emph{Sunflowers}
\end{quoting}

Seven participants described realizing aspects about their work they hadn't realized before (such as how long it takes them to complete part of a painting), or slowing down and making more deliberate creative decisions as a result of writing down their reasoning for each phase of the art-making process. 
Some added that it would be useful to look back at their own processes in the future, saying that it was difficult to remember how their own work began:

\begin{quoting}
\emph{Besides the usefulness of seeing other artists work and a different idea of how they work, [Mosaic] lets you look back at your own and kind of see that, oh, you started that bad. Sometimes you get caught up in that last image and be like, oh, I think it finally came together. It's not too bad. Then you feel like, oh, can I do something like that again? Or you start doing something and it looks horrible, but you don't remember that something else you did looked horrible to begin with.}
---P2
\end{quoting}

%

In other words, Mosaic projects were distinctly unlike traditional tutorials, acting instead more like diaries; they became tools for reflection.

\begin{table}[!t]
\centering
\begin{tabularx}{\columnwidth}{llll} 
\toprule
WIP Type & Count & \% & Example \\
\midrule
Describing action & 177 & 53\% & ``I brush color onto...'' \\
Justifying action & 75 & 23\% & ``I did this because...'' \\
Intent & 37 & 11\% & ``I wanted a feeling...'' \\
Struggle & 23 & 7\% & ``I'm having trouble...'' \\
Assessment & 7 & 2\% & ``I think it looks...'' \\
Idea & 6 & 2\% & ``The idea came from...'' \\
Ask for help & 4 & 1\% & ``Which option is best?'' \\
Goals for growth & 4 & 1\% & ``I wanted to improve...'' \\
\bottomrule
\end{tabularx}
\caption{The types of descriptions written by creators to accompany work-in-progress snapshots.}
\label{tab:wip_categories}
\end{table} 

\begin{table}[!t]
\centering
\begin{tabularx}{\columnwidth}{llll} 
\toprule
Comment Type & Count & \% & Example \\
\midrule
Specific like & 64 & 35\% & ``I like the colors...'' \\
Thanks (creator) & 34 & 18\% & ``Thank you!'' \\
Answer (creator) & 20 & 11\% & ``I did this by...'' \\
Encouragement & 18 & 10\% & ``Looks great!'' \\
Suggestion/critique & 13 & 7\% & ``I would change...'' \\
Intent (creator) & 10 & 5\% & ``I hope it looks...'' \\
Commiseration & 6 & 3\% & ``Painting is hard...'' \\
Technique & 3 & 2\% & ``I'll have to try that...'' \\
Communication & 4 & 2\% & ``Easy to understand...'' \\
Response action & 4 & 2\% & ``I'll make sure to...'' \\
Question & 2 & 1\% & ``How did you...'' \\
Other & 7 & 4\% & \\
\bottomrule
\end{tabularx}
\caption{The type of comments written about projects.}
\label{tab:comment_categories}
\end{table} 

\subsection{Feedback helps validate process}
Four participants stated that feedback was difficult to get in existing art communities, attributing this to the audience that was drawn by a platform oriented around gaining exposure. The relationship they had built up with others on these sites were more of a ``celebrity-fan'' relationship, rather than a relationship between artists who can help each other:

\begin{quoting}
\emph{Most of the comments on [DeviantArt] are ``Oh, I love it, amazing.'' Which is great, I'm always grateful that people like my work, but if you're looking for anything specific you're not going to get it there.}
---P3
\end{quoting}

For this reason, six artists stated their current preferred method for receiving feedback was to ask artists they know in real life or friends they trusted, but this can quickly exhaust social capital.

Mosaic, on the other hand, was described by seven participants as a very artist-centric platform. 
Users made a total of 153 comments, with each project receiving an average of 2.22 ($SD=2.19$) comments and with users making an average of 3.85 comments ($SD = 6.08$) each. 
The median time for comments to appear after a user made an update to a project was 16.52 hours, and projects received an average of 0.61 total comments ($SD = 1.32$) prior to its last update.
Only 10\% of comments were the simple encouragement typically seen in existing online art communities, with most other comments remarking on specific aspects of the process described by the creator, commiserating with the creator about the difficulty of the process, asking questions, or providing suggestions or critique (Table~\ref{tab:comment_categories}). By aiding artists in revealing the process behind an artwork, Mosaic reinforced a social norm of writing specific, relevant feedback:

\begin{quoting}
\emph{[Comments on Mosaic are]...if they say they like something, they seem to actually say what about it they liked... They seem a little bit more... informed as fellow artists. It's not just ``Oh, that's cute,'' or ``That's pretty.''} ---P2
\end{quoting}

Artists were very open to both negative and positive feedback. When asked about the kind of comments they wished they could get more often, five participants said they wanted feedback not to neccessarily to improve their work, but to validate whether or not their creative intent was coming through in their output and to see if they were on the right track with their progress.
More generally, participants described good feedback as specific, timely (that is, received during the creative process rather than afterwards), and relevant to current goals (rather than suggesting other goals); participants stated they would ignore feedback that was contrary to their creative intent. Mosaic seemed to allow other creators to pinpoint the intent of the creator posting artwork (often because the creator now had the opportunity to explain their goals and reasoning through works-in-progress), leading to more informed feedback from the community as a whole and allowing creators to use other users as a mirror to help them reflect on whether or not they were able to achieve their goals.

\subsection{Teaching through WIPs, teaching through feedback}

Creators approached uploading and composing works-in-progress on Mosaic as an informal teaching opportunity, with eight participants describing their imagined audience \cite{marwick2011tweet} as other artists at a skill level just below their own or even to ``a me from the past'' (P5). For the most part, artists documented how they overcame some struggle or achieved some goal, and described their project in terms of teaching what was learned to others.
Occasionally, if they found themselves stuck on solving a problem, they would break from this teaching role and ask for help. Overall, however, participants felt that each of their works-in-progress needed to represent substantial progress on the project so that they would have something to say to their audience.
Participants were aware of the value of clearly communicating their process, with eight participants saying that posting works-in-progress was only useful when presented in chunks that made sense or when it was complete:

\begin{quoting}
\emph{I actually discarded a few [works-in-progress] that didn't seem like they made much of a difference in between. I just kind of chose some of the biggest ones you could see. I added this detail or changed colors or added more facial details here. Anything you could see an actual progress to.}
---P2
\end{quoting}
This reflects votes from other Mosaic users during the peer voting round to decide which projects would be included in the zine. 
Voting was open to all Mosaic users; out of 46 registered users, 33 users participated in the peer voting round by selecting the three Mosaic projects they thought should be included in the zine. A Poisson regression showed that projects with the most votes were those which had more works-in-progress ($\beta=0.114, p < 0.01$), as well as more comments ($\beta=0.135, p < 0.01$).

Paradoxically, Mosaic users seemed to approach writing feedback as a teaching opportunity as well; five participants mentioned that being able to view other people's works-in-progress influenced their motivation to write feedback:

\begin{quoting}
\emph{If someone is posting as they work on their work, you actually feel like you're right there encouraging them if you're giving them feedback through steps and everything... If they post all their updates and they post when they're finished, you actually feel a connection because you felt like you were cheering them on the entire time they were working on this thing. Now you see the finished the product and you're like, ``Dude, that's awesome.''} 
---P8
\end{quoting}

This may be explained by an underlying ideal of fairness explicitly mentioned by two participants:

\begin{quoting}
\emph{I do think that part of what's interesting about this site is that you do get to see all these middle steps. It seems a little unfair to not share that if I'm taking that in from other people.}
---P7
\end{quoting}

\begin{quoting}
\emph{I think it's the point of the community, in part, is to give and take critique and I think that's really cool.}
---P6
\end{quoting}

That is, posting about creative process actually afforded reciprocal give and take in a creative community; creators ended up framing their projects as gifts of knowledge to others, but this was also the same mechanism through which people received feedback and help.

\subsection{Showcasing failure is uncomfortable}
The fact that existing communities focus on sharing final outcomes also means that first impressions are important on these websites, further discouraging sharing in-progress work. It is often difficult to bring viewers back to see updates to a creative piece in progress:

\begin{quoting}
\emph{I normally try to upload everything... when I'm done... that way, if people are viewing it, they're not like, ``Oh, this is cool but I don't know where it's going. I don't want to come back and look at this,'' or they'll forget about it.}
---P4
\end{quoting}

In addition, six participants mentioned they experience apprehension when sharing their work online due to a lack of confidence in their skills or negative experiences with aggressive commenters from the past.
However, six artists also mentioned documenting their process in a community allowed them to contrast their process against others, creating an environment where sharing process was normal and easing fears about posting content:

\begin{quoting}
\emph{[Sharing process] encourages people to share what they know... they don't feel like they're in direct competition because we're all learning at the exact same time, just at different paces.}
---P8
\end{quoting}

In other words, posting content on Mosaic became less about trying to prove worth to an audience and more about the journey of each individual creator.

\section{Discussion}
Through our evaluation of user activity on Mosaic, we learned how a community of sharing works-in-progress can help creators give and receive specific feedback and reflect on their creative practices. Interestingly, we also discovered how creating an environment that encourages sharing process can also help creators feel comfortable sharing unfinished work or asking for help. In this section, we generalize our findings by discussing design implications for future social computing systems and proposing a design space for creativity support tools that encourage useful creative outcomes---such as mistakes, failures, prototypes, and experiments---beyond traditional notions of success.

\begin{figure}[!t]
\centering
\includegraphics[width=1\columnwidth]{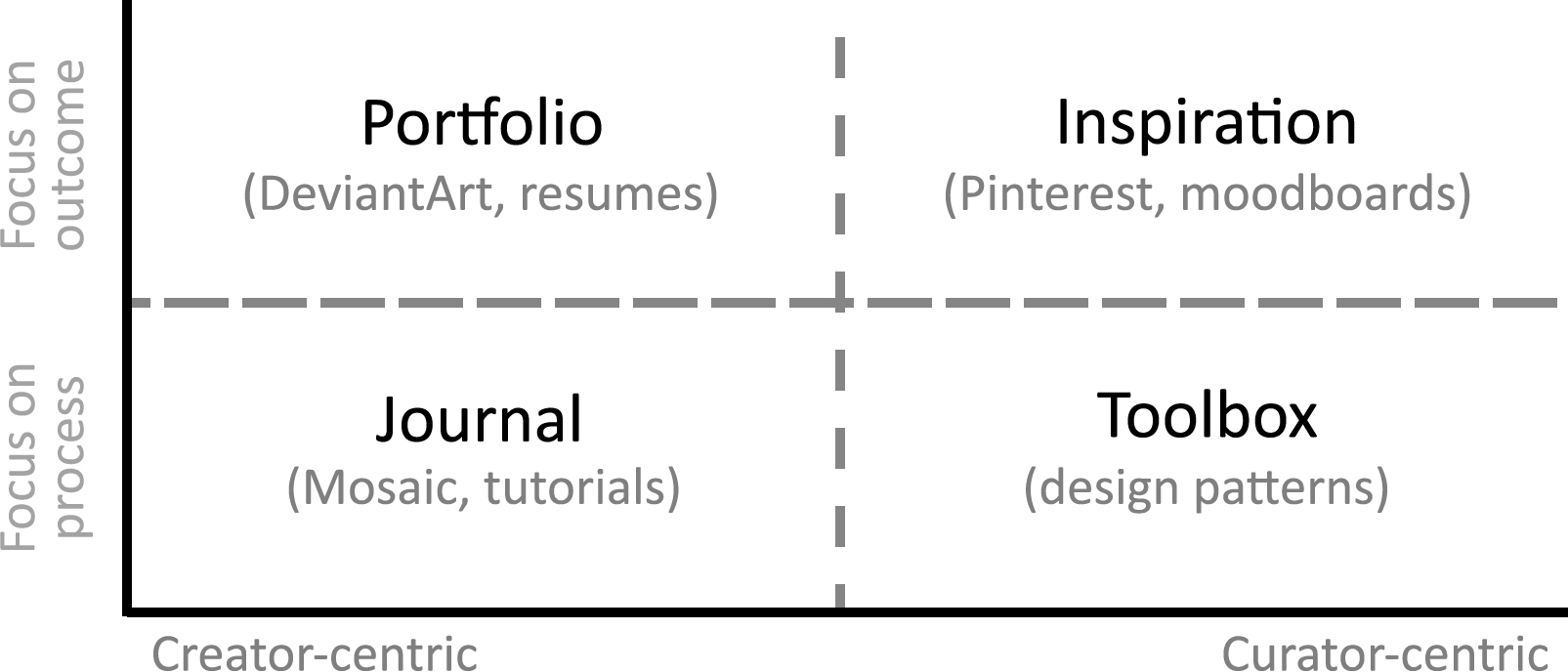}
\caption{A design space of possible creative communities based on what creative outcomes are being shared and by who.}
\label{fig:designspace}
\end{figure}
 
\subsection{A design space for sharing creative work}
Though Mosaic focused primarily on supporting painting and illustration, its interface for sharing snapshots of in-progress work over time could apply directly to several other domains, including music, writing, and design. This would likely work best for domains where there is a single artifact that represents the whole creative work. Something like film-making consists of writing a script, a casting process, days of actual filming, and more; it is hard to say what a snapshot of work would look like in this case. In addition, Mosaic may best benefit work with a smaller scope; it's easier to share and give feedback on snapshots of a short story, for example, than an entire novel.

At a high level, however, Mosaic expands the space of possible designs for creative communities by broadening the scope of useful artifacts that creators may want to share---namely, creative processes. 
Doing so increases social translucence~\cite{Erickson:2000:STA:344949.345004}; social interactions are no longer solely based on the final result but also on an awareness of what a creator has done to the create the work and why certain creative choices were made.

In Mosaic, we saw evidence that the design of a creative community can affect users' views on what type of content is valuable and useful for the rest of the community. Many existing creative communities are \emph{creator-centric}, meaning that they focus primarily on allowing creators to share their own content in publicly viewable portfolios. However, communities can also be \emph{curator-centric}, and focus more on allowing creators (or fans) to showcase or curate the work of others; Pinterest~\cite{pinterest}, where users can gather content in themed ``boards,'' is one example of a community designed around social content curation. These design attitudes are not mutually exclusive, as many communities (including DeviantArt) also support curation by allowing users to favorite work by others and organize and share these favorites.

These two axes---whether users are sharing their own or curating others' content, and whether users are focusing on sharing outcomes or process---reveal a design space for online creative communities for sharing work (Figure~\ref{fig:designspace}). The upper-left quadrant contains communities for creators to showcase the outcomes of their own work; the upper-right quadrant contains communities for creators to curate work of others which is commonly used to support creative activities such as collecting inspiration and examples. The lower-left quadrant represents communities that focus on enabling creators to share their own creative processes with each other, and includes communities like Mosaic. The lower-right quadrant represents an open design opportunity: communities that allow creators to curate the creative processes of others. This could simply be the curation and collection of tutorials, or one could imagine a community where creators create customized libraries of socially vetted techniques.

This design space is certainly not comprehensive, but acts as a starting point for thinking about the design of creative communities in a broader way. One could imagine, for example, inverting this design space to focus on negative outcomes rather than positive ones: communities that share portfolios of good work become spaces for creators to share their worst failures; creators could collect examples of finished work they do not like to help them scope the range of possible ideas for their next art piece; and creators could even document creative processes that ended in failure (i.e., what not to do), which would be valuable from a learning perspective for both themselves and for others.

\subsection{Process as intellectual property}
Creators who share their work online often worry about \emph{art theft}, where someone reposts their work somewhere else without credit (or while claiming to be the creator). In our formative study, we asked interview participants whether art theft was a concern when posting works-in-progress. Surprisingly, participants stated the consequences of art theft were (while annoying) mostly harmless, and doubted that someone would take the effort to steal their work since they felt they were not particularly famous.
This response is likely due to how we recruited participants, since we expressly sought creators with similar backgrounds and skills to make it possible for creators to give feedback to each other.
Potential theft may be a more pressing issue for those who consider process part of their intellectual property (indeed, a process can be considered a type of patentable invention). While this might suggest that Mosaic's focus on sharing process is not applicable to domains where sharing early ideas may result in loss of competitive advantage (e.g., startups), one could imagine using a system like Mosaic internally to facilitate transparency and feedback within a team.

\subsection{Growing the Mosaic community}
How does a system like Mosaic grow? 
Maintaining communities like Mosaic can be difficult, as shown by the closure of popular communities that have attempted to focus on process but have transitioned back into showcasing outcomes~\cite{behanceclosed}.
Mosaic was described by interview participants as making it much easier to upload series of works-in-progress compared to existing painting and illustration sites, and some participants even reported sharing links to their Mosaic projects on their other social media profiles. However, while we were able to find positive effects of Mosaic's design among a community of users who were already actively using Mosaic, observing less active users or people external to Mosaic would give us a better sense of why people do (or do not) participate actively in the Mosaic community or the value they derive from visiting Mosaic as a lurker. For example, is it unrealistic to expect that most creators will take the time to post detailed works-in-progress? Or, do people find they enjoy viewing Mosaic projects without community interaction? In other words, what would the social landscape of something like Mosaic look like at larger scale?

For example, though Mosaic users were able to give specific feedback to one another, there was a large variance in the average number of comments written by users who had uploaded at least one project. It may have been difficult for some users to find projects to give feedback on, as Mosaic's main method to display projects was to display a feed of recent activity from the community as a whole. This problem would only become larger as the community grows. However, an increase in community size may help projects receive more feedback in a timely manner (that is, while the project is still in progress).
It may be worth expanding on Mosaic's feature of being able to flag critiques and use this as a signal to push projects wanting critique to other users, or even create a matchmaking service that connects project wanting feedback with users who upload similar projects. As another example, while 35\% of comments on Mosaic were specific feedback about what the commenter liked about a creator's work, other types of useful comments such as critiques (7\%) and comments on techniques used (2\%) were less common. Approaches such as structuring feedback using guidelines or templates~\cite{Xu:2014:VGS:2531602.2531604} may further support creators in writing specific and timely feedback for one another.

In addition, how does Mosaic maintain its focus on process as it grows? 
This paper focused encouraging community contribution by enabling creators to share their process, but healthy communities also require committed users, regulation of behavior, and procedures for attracting and socializing new members~\cite{kraut2012building}.
The creators that participated in our study were already familiar with existing art communities and had established art practices; it would be interesting to see how users new to art communities are affected by Mosaic's social norms (as well as how the community reacts when these norms are violated by new users).
One possibility is that Mosaic's decision to structure content creation in terms of snapshots of progress will help convey community values. For example, this may encourage new users to describe their work in detail and make it easier for others to identify and act on opportunities to help~\cite{Teo:2014:FFF:2531602.2531731}.
In addition, we found that Mosaic users approach posting works-in-progress as teaching opportunities; these works-in-progress may thus also act as proof that the artist put a nontrivial amount of effort towards their work. Other users may take this as a signal that the artist will spend a similar amount of effort incorporating any help they receive.
In future work, it would be interesting to study a community composed of creators unfamiliar with online art communities and see if Mosaic's design ``autocorrects'' the behavior of users who are unfamiliar with (or ignore) the social ideals of exquivalent exchange expressed by some of our interview participants.

\section{Conclusion}
In this paper, we make two major contributions. First, we demonstrate the potential benefits of an online creative community based around sharing works-in-progress creations. We did this by building Mosaic, an online creative community, and conducting an observational study where we interviewed creators about their interactions with each other as well as their artmaking process. Artists described being able to give and receive more helpful feedback on their work and feeling more comfortable sharing unfinished work and mistakes (relative to other creative communities). Second, we generalize the approach we proposed through Mosaic and generate a design space demonstrating opportunities for new types of creative communities by expanding our idea of useful creative outcomes.
While not comprehensive, the examples discussed here illustrate the possible ways we can fill the gaps in support for communities of creators left by current systems. By explicitly designing to create space for exploration, process, and failure in creative tools and communities, we may better enable creators to not just achieve but also grow.

\section{Acknowledgments}
We would like to thank our colleagues that helped shape this research with their valuable feedback. We also thank all the artists who participated in this study for their time and expertise. This material is based upon work supported by the NSF GRFP under Grant No. DGE-114747 and by the Hasso Plattner Institute Design Thinking Research Program.

%
%
%
%
%
\balance{}

\bibliographystyle{SIGCHI-Reference-Format}
\bibliography{sample}

\end{document}